# Link Quality Aware Channel Allocation for Multichannel Body Sensor Networks

Weifeng Gao[a], Zhiwei Zhao[*,a], Geyong Min[b], Yue Cao[d], Hancong Duan[a], Lu Liu[c], Yimiao Long[a], Guangqiang Yin[a]

**Abstract**

Body Sensor Network (BSN) is a typical Internet-of-Things (IoT) application for personalized health care. It consists of economically powered, wireless and implanted medical monitoring sensor nodes, which are designed to continually collect the medical information of the target patients. Multichannel is often used in BSNs to reduce the spectrum competition of the tremendous sensor nodes and the problem of channel assignment has attracted much research attention. The health sensing data in BSNs is often required to be delivered to a sink node (or server) before a certain deadline for real time monitoring or health emergency alarm. Therefore, deadline is of significant importance for multichannel allocation and scheduling. The existing works, though designed to meet the deadline, often overlook the impact of the unreliable wireless links. As a result, the health sensing data can still be overdue because of the scheduled lossy links. Besides, potential collisions in the schedules also incur considerable delay in delivering the sensing data. In this paper, we propose a novel deadline-driven *L*ink quality *A*ware *C*hannel *A*ssignment scheme (LACA), where link quality, deadlines and collisions are jointly considered. LACA prioritizes links with urgent deadlines and heavy collisions. Besides, LACA allows the exploition of the spare slots for retransmissions on lossy links, which can further reduce the retransmission delay. Extensive simulation experiments show that compared to the existing approaches, LACA can better utilize the wireless spectrum and achieve higher packet delivery ratio before the deadline.

*Keywords:* channel assignment, multichannel, body sensor network, link quality aware, deadline driven

# 1. Introduction

Interenet of things (IoT) enabled healthcare is a promising research direction in recent years [1, 2]. Body sensor network (BSN) is a novel IoT enabled healthcare application specified for personalized health care and has attracted much research attention in recent years [3, 4, 5, 6]. A BSN often consists of a number of sensor nodes attached to or implanted in the target patients, which continually collects the health information and transmit it to the data sink for personal health analysis [7]. As the microelectronic technology develops, more and more types of wearable sensor nodes are designed and applied in BSNs [8, 9, 10, 11]. For example, the smart wrist rings [12] can be used to measure the physiological parameters such as heart rate, blood pressure, skin temperature. The smart shirt [10] can be used to measure the electrocardiogram (ECG), photoplethysmograph (PPG), heart rate, blood pressure, body temperature.

These devices, while collecting more comprehensive human health information, have intensified the wireless spectrum competition as most of them work in the same frequency. Therefore, multi-channel is employed as an important means to reduce the collisions and better utilize the wireless spectrum resources [13].

In typical multi-channel BSNs, each node works in a duty cycle and the network operates as the duty cycle repeats. Channel allocation is a fundamental and significant problem. Each link is assigned a slot in the duty cycle and a corresponding communication channel in the slot. Some channel allocation schemes have been proposed in recent studies. Jun et al. [14] exploit the capture effect to reuse the good channels in channel allocation, which can improve the transmission efficiency. Wang et al. [15] propose a decentralized scheduling algorithm to allocate the channels and time slots in order to jointly reduce control overhead, packet transmission delay and duty cycle. Bui et al. [16] provide soft real-time and bandwidth guarantees by avoiding collisions with multiple channels using real-time chain. Saifulah et al. [17] aim at establishing real-time channel scheduling in Wireless HART network.

Since health data is important for real time monitoring [10] and health emergency alarm [12], the health data is often required to be delivered to the sink before given deadlines. However, most of the existing works cannot well support the deadline-driven data routing and transmissions in BSNs. The reason is two-fold. First, the lossy nature of wireless links is not carefully considered in the channel assignment and scheduling. The existing works often assume a packet can be successfully delivered in one single slot. However, this is not true for real world scenarios where wireless links can be easily interfered [18, 19]. As a result, the arrival delay of the data packets are often under-estimated and the resulting schedules may lead to data overdue problem. Second, retransmissions are not considered in the existing works, while there are often a number of slots are left unused. In the existing works, the retransmission is typically postponed by an entire duty cycle [20, 21], which adds large delay to the end-to-end delivery time.

To address the above limitations, we propose a novel deadline-driven *L*ink quality *A*ware, deadline-driven *C*hannel *A*llocation scheme (LACA) aiming to maximize the packet delivery ratio (PDR) before deadline as well as minimize the wireless collisions.

Compared with the existing works, LACA has two salient features: *First*, LACA jointly considers link quality, wireless collisions and deadlines in the scheduling. A novel metric that takes both deadline and collisions is proposed to prioritize the paths with more urgent deadlines and heavier collisions. For assigning links for a specific path, channels with better link quality and less collisions are preferred. The links that have more collisions are more likely to be assigned with better channels/slots. *Second*, instead of trying to minimize the assigned slots, we further assign "backup" channels/slots to the lossy links. With these "backup" slots, the lost packets can be intermediately transmitted and the delay can be significantly reduced. It is worth noting that these "backup" slots do not incur extra energy overhead since they are activated only when packet losses happen.

We implement LACA in simulation experiments and compare its performance with the existing works extensively. The evaluation results show that LACA greatly improves the PDR before deadline in multichannel BSNs and does not incur extra energy overhead.

The major contributions of this paper are listed as follows.

1. A novel metric for channel allocation is proposed, which prioritizes the paths with more urgent deadlines and heavier inter-path collisions.
2. A novel link quality aware channel allocation mechanism (LACA) is proposed, which jointly considers link quality, collisions and possible retransmissions.
3. We evaluate LACA's performance by extensive simulation experiments and the results show that LACA significantly improves the packet delivery ratio (PDR) before deadline than the existing works.

The rest of this paper is organized as follows. Section 2 presents the related works and the motivation of our work with illustrative examples. Section 3 presents the design details of LACA. Section 4 evaluates LACA with simulation experiments. Section 5 concludes this work.

## 2. Related Work and Motivation

The health data in BSNs is often required to be delivered to the sink before a given deadline in order to achieve real-time patients monitoring or health alarms. Therefore, deadline is a more critical issue in BSNs than that in general sensor networks.

Several studies have been working on the problem of deadline guaranteed channel scheduling. These works could be divided into two categories according to the channel diversity: 1) works using single channel scheduling. 2) works using multichannel allocation and scheduling.

### 2.1. Works using single channel scheduling

Works on single channel mainly include [22, 23, 24, 25, 26]. These works focus on scheduling the links to guarantee the deadline in single-channel networks. Kumar et al. [22] try to decrease the distance among the source and destination by using multiple short-distance hops instead of long-distance hops. Li et al. [23] prove that providing deadline guarantee for message transmissions in wireless sensor networks is NP-hard. They propose a scheduling scheme with MAC layer back-offs and the spatial reuse of the wireless channel. Shakkottai et al. [24] study the problem of scheduling transmissions with deadlines over one channel, gathering the number of packets lost due to deadline expiry. The work in [25] considers both deadline and link quality into account, they use multiple path routing to guarantee the reliability. Andrews et al. [26] jointly consider the end-to-end delay constraints and throughput requirements. To summarize, these works try to find the best schedules of the colliding links to reduce the end-to-end transmission delay.

*2.2. Works using multiple channels*

Works on multiple channels mainly include [16, 17, 15]. These works use channel diversity to improve the performance of scheduling. Bui et al. [16] provide soft real-time and bandwidth guarantees by avoiding collisions with multiple channels using real-time chain. Saifulah et al. [17] aim at channel scheduling with deadline in Wireless HART networks, where the wireless links are assumed to be highly reliable. Wang et al. [15] propose a decentralized scheduling algorithm to allocate the channels and time slots in order to jointly reduce control overhead, packet transmission delay and duty cycle. In these works, the underlying assumption is that a packet transmission is successful in one slot and the corresponding channel.

It is worth noting that BSNs are often deployed in indoor environment, where there may be more severe interference such as WiFi interference. As a result, channel allocation and scheduling in BSNs has its distinct features:

1. Some channels are overlapped with WiFi channels and may not be eligible for efficient transmissions, which leads to less channels available for channel allocation compared to the typical outdoor sensor networks.
2. Due to the interference, the link quality in BSNs is more time-varying and can greatly affect the network performance.

Therefore, link quality should be taken in the deadline guaranteed channel allocation.

Our proposed work in this paper has two important differences with the above literature. First, LACA jointly considers collisions and deadline constraints to prioritize the path links with more urgent deadlines and heavier collisions, which are more likely to be assigned with better quality channel/slot pairs. Second, link quality is considered in the channel/slot allocation. This can yield more reliable scheduling than the works that assume packets can be successfully delivered in one single slot.

There are also channel scheduling works [27, 28] that consider link quality. However, they are not designed for deadline-guaranteed data transmissions.

*2.3. Motivation*

Next we present the motivation of our work using two illustrative examples.

*2.3.1. The impact of wireless lossy nature*

Figure 1 shows an example where nodes A-F intend to transmit health data packets to the sink node G. The length of the duty cycle is five, i.e., each node repeats the schedule every five slots. The small rectangles beside each node denote the data arrival time (left) and the deadline (right). For example, node B receives (or generates) the data in slot 1 (the left rectangle) and should deliver the data to node G no longer than three slots (the right rectangle). Using the channel allocation scheme in [17], the channels and slots are sequentially assigned to each path and the allocation result is shown in Figure 2(a). With the allocation, each node can deliver the generated data to node G if there are no packet losses. However, if packet losses happen, say, packet transmission from A to F fails, another duty cycle of five slots will be incurred and the packet cannot reach node G before deadline. At the same time, there are many spare slots that are not assigned any channels or links. Intuitively, each source node 1) assigns better-quality channels and slots to the important links and 2) could start retransmission right in the next slot after the packet is lost. With retransmissions, the data packets can still be delivered to node G before the deadline (e.g., node A can still deliver packets to G within four slots). We manually assign channels/slots with good quality to the links and also assign some spare slots/channels to the lossy links (with low link quality). The new channel assignment is shown in Figure 2(b). Some links are assigned multiple slots, of which some (slot 0 with channel 0 and 1, slot 1 with channel 2, slot 2 with channel 1, slot 3 with channel 0 and slot 4 with channel 2) are for the first time transmissions and the others (slot 1 with channel 1, slot 2 with channel 2, slot 3 with channel 1) are for the possible retransmissions when packet losses happen. Please note that link "A-F" cannot be retransmitted on slot 1 with channel 2 because link "B-F" has already been allocated with it. These two links will conflict with each other if they transmit in the same time. Comparing slot 1 with channel 0 and slot 2 with channel 2, the latter is with better link quality and is then assigned for "A-F"'s retransmission. Since the links are activated in the corresponding retransmission slots only when the node detects packet losses, these retransmission slots will not incur extra energy overhead when there are no packet losses. On the other hand, when there are packet losses, these slots can be used to further reduce the transmission time. In our example, node A's expected PDR with the new allocation (Figure 2(b)), which is higher than the PDR of the original allocation where link quality or retransmissions are not considered.

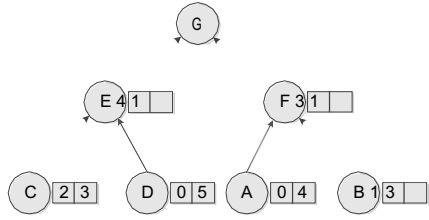

Figure 1: Motivating example for the impact of lossy links.

### 2.3.2. Tradeoff between deadlines and collisions

For assigning the links with deadlines, the existing works often follow the "Urgent First" rule: the path with the most urgent deadline should be assigned first. Following the rule, we assign the channels/slots for a small network shown in Figure 3. According

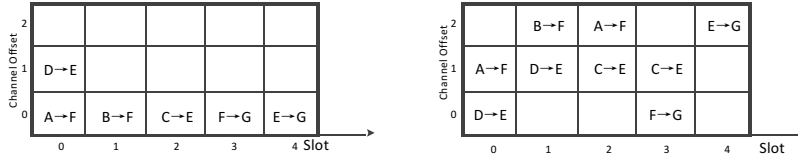

(a) Sequential channel allocation wihtout retrans- (b) Channel allocation with link quality and misssion.    retransmission.

Figure 2: Channel allocation comparison.

to [23], the urgency is defined as:

$$d = dl - pl \qquad (1)$$

where $dl$ denotes the deadline of the slots and $pl$ denotes the path length from the current node to the sink node. For example of node D, $d(D) = 2-2 = 0$. Using the above metric, the assigning order of all paths is: "F-G","B-F-G","D-E-G","E-G","C-E-G" and "A-D-E-G". Then paths "F-G", "B-F-G", "D-E-G", "E-G" and "C-E-G" are sequentially assigned with the slots/channels as shown in Figure 4(a). Finally when assigning path "A-D-E-G", link "A-D" should be assigned either in slot 0 or slot 1 in order to reach the deadline. However, there is no available slot/channel for "A-D" because both slot 0 and slot 1 with the two channels are already occupied by the priorly assigned links. As a result, there is no available channel/slot for link "A-D" to deliver the data from node A to node G before deadline. To address this problem, deadline and collisions should also be taken into consideration in assigning the links at the same time. We re-order the assigning sequence and the assignment is shown in Figure 4(b). We can see that all links are assigned appropriate slots and channels such that the deadline can be met. Specifically, the data from all nodes can be delivered to node G before deadline and the colliding links are not assigned in the same channel and slot.

Our key idea is to design a novel metric to sort the paths needed to be assigned. With the sort using the metric, the number of nodes that can reach the deadline can increase and the collisions can be reduced at the same time.

From the above two illustrative examples, it can be revealed:

1. Link quality can significantly impact the channel allocation. The potential packet losses can greatly add to the packet delivery time, which is critical especially for deadline-driven networks such as BSNs. The "additional" slots, which are often left spare by the existing works, can be used for adaptive retransmissions and should be considered in the channel allocation.

2. The sequence of the assigning paths can greatly affect the channel allocation. If the path sequence is not carefully designated, some colliding links may not be able to be arranged in the channel schedule and thus the corresponding deadlines cannot be met.

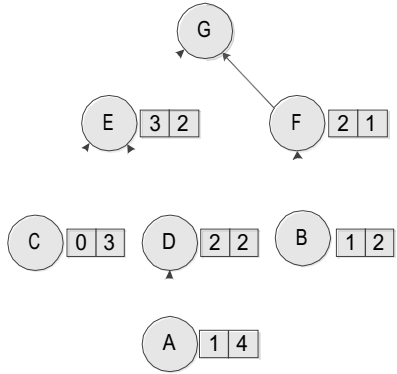

Figure 3: The example showing the tradeoff between deadline and collisions.

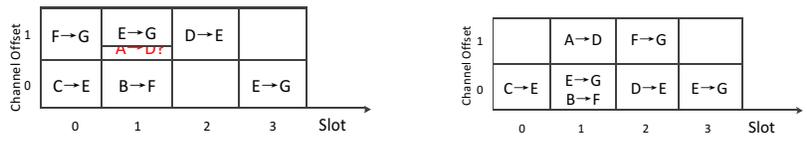

(a) Urgent paths first allocation.

(b) The ideal allocation which meet the deadline requirement and avoid collisions.

Figure 4: Path sequence comparison for channel/slots allocation.

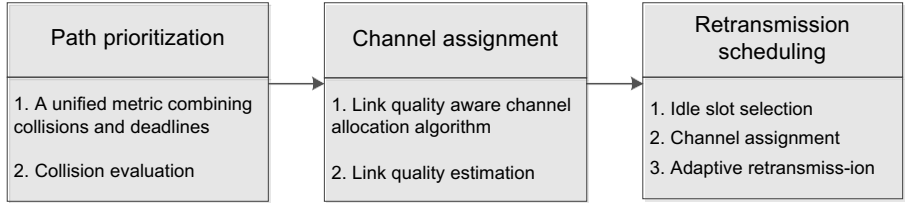

Figure 5: LACA Overview: Paths are first prioritized before allocation; Then channels and slots are assigned to the paths by order; Retransmissions are scheduled after channel allocation.

To address the above two problems, A promising alternative to first prioritize the paths that have more urgent deadlines and heavier collisions. Once such paths are assigned with specific channels and slots, other colliding paths can be more easily assigned in the schedule. Then we consider both link quality and retransmissions in the channel allocation.

## 3. Main Design of LACA

In this section, we present the main design of deadline-driven *L*ink quality *A*ware *C*hannel *A*llocation (LACA). In LACA, the path links to be assigned are first sorted with the proposed metric, which prioritizes the links that have more urgent deadlines and heavier collisions using the proposed metric in Section 3.1. For each path, we assign the links sequentially from the link directly connected to the sink node to the link starting from the source node as described in Section 3.2. For each link to be assigned, the channels/slots with better link quality are first selected for assignment. When all paths are assigned and there are still some slots that are not assigned to any links, we arrange to assign these spare slots for possible future retransmissions. During the channel allocation process, link quality is estimated using the in-packet corruptions [29, 30, 31] for each pair of channel and slot. The link estimation is described in Section 3.3. The notations used throughout this paper is summarized in Table 1.

### 3.1. Path assignment prioritization

Given *n* available channels, LACA tends to assign time slots and channels to the appropriate links to guarantee the deadline of data transmission on each path. To guarantee the deadlines on paths, the algorithm iteratively processes one path at a time until all paths are assigned. Since data generated from every node carries its deadline $D_i$, which limits the allocation to the destination node, paths with lower value $d_i$ (which equals the deadline minus the effective length of the path from the source node to the destination node) tend to be more urgent to be transmitted. To guarantee the deadline requirement, these paths should be prioritized. At the same time, as analyzed in Section 2.3, links with heavier collisions should also be prioritized as well because the links left are non-colliding and have more choices on the available slot/channel pairs.

Table 1: Notations

| Notation | Description |
|---|---|
| $s_i$ | The source node $i$ |
| $g_i$ | The data generation time for $s_i$ |
| $d_i$ | The deadline for the data of node $s_i$ |
| $p_i$ | The path from $s_i$ to the sink node |
| $l_i$ | The link $i$ |
| $q_{ij}^m$ | The link quality of link $m$ in channel $i$ and slot $j$ |
| $S$ | The set of source nodes, $S = \{s_i\}$ |
| $G$ | The set of data generation time for each node, $G = \{g_i\}$ |
| $D$ | The set of deadlines for each source node, $D = \{d_i\}$ |
| $P$ | The set of paths from the source nodes to the sink node, $P = \{p_i\}$ |
| $L_i$ | The links in path $i$, $L_i = \{l_j \in p_i\}$ |

To achieve the above prioritization, we propose a novel metric as follows.

$$m_i = \alpha d_i + (1 - \alpha) c_i \quad (2)$$

where $d_i$ denotes the urgency and $c_i$ denotes the collisions. $\alpha$ is a weighting parameter, where a larger $\alpha$ prioritizes the urgent paths more while a smaller $\alpha$ prioritizes the heavier-collision paths more. The above two factors are considered important due to that: 1) BSNs are often deployed in indoor environments where the network is dense. 2) The data often matters with the patients health status, which is critical for real-time monitoring and health alarms. The parameter $\alpha$ can be tuned according to the practical deployment of the BSNs. If the network is not dense, $\alpha$ can be set large to weigh more on the urgency. Otherwise if the network is very dense and is used for general patient monitoring without real-time requirement, $\alpha$ can be set small to weigh more on the collisions. The urgency $d_i$ is calculated as:

$$d_i = dl_i - pl_i \quad (3)$$

The collisions $c_i$ is calculated as the number of colliding paths of path $i$. $c_i$ is obtained using the algorithm Alg. 1, where every colliding link of path $i$ is accounted in $c_i$.

*3.2. Channel allocation*

After the paths are sorted, we start the link quality aware channel allocation. The link quality is obtained using the scheme described in Section 3.3. When scheduling a path, we first obtain the available channels and the link qualities of corresponding slots and channels. Then we choose $r$ time slots with best $r$ quality channels, where $r$ is the number of links in the path and each time slot is for one link. After that, we start from the first link in the path and assign the best quality slot in the available time slots. Then we assign the second link with the second best slot/channel pair. The above process repeats until all links on all paths are assigned with certain channel and slots pairs.

**Algorithm 1** Conflict count
___
**Input**: Path starting from a source node $s_i$, $p_i$;
The set of links in $path_i$, $L_i$;
The set of all links in the network, $L$;
**Output**: $p_i.cnum$, the number of conflict links with $path_i$;
**for** *each $l_j \in L$* **do**
    $l_j.cnum = 0$; //cnum denotes the number of colliding links
    **for** *each $l_k \in L$* **do**
        **if** *$l_j$ collide with $l_k$* **then**
            $l_j.cnum+ = 1$;
        **end**
    **end**
**end**
**for** *each $l_n \in L_i$* **do**
    $p_i.cnum+ = l_n.cnum$;
**end**
return $p_i.cnum$;

After all links are assigned with slots and channels, we further check whether there are unused channel/slot pairs with link quality above a threshold $T_q$, where $T_q$ can be set according to the average link quality in the network. If there are such channel/slot pairs, we further exploit these pairs for the retransmission on links with the worst link quality for possible retransmissions. There are two impact factors in assigning the retransmission slots. 1) These channel/slot pairs should be of good quality. 2) These channel/slot pairs should be available for the successive slots for the target links, such that retransmissions can be done in these slots. Otherwise, at least a duty cycle of the slots will be required and the retransmission slots cannot be effectively employed to reduce the delivery delay.

*3.3. Link quality estimation*

Our allocation scheme relies on the link quality for each channel/slot pair. However, one challenge is to obtain the real time link quality profile for each channel/slot pair as only one channel's quality can be measured at a time. Similar with [31, 30], we use the 32KHz timer to measure the in-packet byte-level RSSI values. The smallest in-packet RSSI value is used as the RSSI base value. Then the correspondence relationship between byte error rate and the RSSI distance from the RSSI base is used to estimate link quality. We measure thousands of packets and collect the RSSI traces to obtain the above relationship, which is shown in Figure 6. The error rate increases when RSSI distance becomes larger. When RSSI distance is larger than 5dBm, the byte error rate is consistently above 90%. With this relationship, we are able to accurately estimate the byte error rate especially when the RSSI difference is large.

With the byte error rate, we can further calculate the link quality as follows:

$$q = \prod_{i=1}^{n}(1 - b_i) \qquad (4)$$

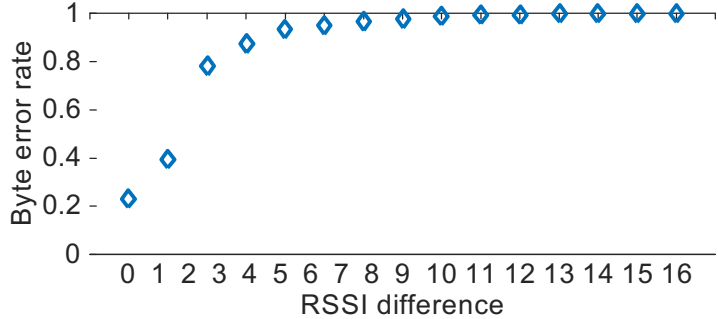

Figure 6: The relationship between byte error rate and RSSI distance.

where $q$ denotes the link quality and $b_i$ denotes the byte error rate of the $i$th byte. Each node periodically exchanges beacons with neighboring nodes in different channels and estimates the link quality corresponding to certain slots and channels. Since the RSSI-based link quality estimation can be done using one single packet, it allows LACA to update the link quality periodically without too much transmission overhead.

*3.4. Retransmission scheduling*

After all the paths are assigned with specific channels and slots, there may be still many slots that are not assigned to any links. In LACA, we allow the use of these slots and channels to further improve the packet transmission efficiency for certain links.

For a link $i$ in one path, the available retransmissions slots include the slots from slot $t_i$ of link $i$ to slot $t_j$ of its following link $j$. If in any slot from $t_i + 1$ to $t_j - 1$ with any channel, link $i$ does not conflict with the already assigned slots, the slot and the corresponding channel with the best quality is assigned to link $i$ as the retransmission slot/channel. The channel/slot pairs with quality below $T_q$ is filtered out before the retransmission assignment. By exploiting these idle time slots and channels, the packet delivery ratio can be further improved.

*3.5. Discussion on further optimizations for LACA*

- The current design of LACA is centralized, where the information from all network links is required. The computational overhead and the information collection incur considerable extra overhead. Besides for the BSNs with highly time-varying links, the assignment results may not be appropriate for the changed link and channel conditions. We will extend LACA as a distributed algorithm to reduce the complexity while achieving the high PDR before deadline.

- The current LACA algorithm assigns the best channel/slot pairs in each round of assignment, which does not achieve the global optimum. In order to achieve global optimum, a possible approach is to formulate it as an integer linear programming problem. However, the challenge lies in that we need to traverse all pairs of channel/slot and paths, which might be of high complexity. We plan to address the challenge and work on the global optimum in our future work.

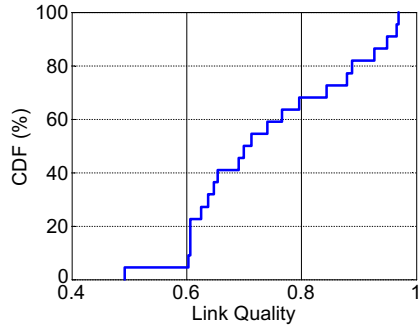

Figure 7: The link quality distribution of the simulated network.

- Mobility of the body sensor nodes can affect the channel allocation because both link quality and the topology might be changed due to the mobility. To deal with the problem of mobility, we can periodically update the topology and link information, and re-schedule the link-channel pairs if the link features are changed. When the path is changed, the allocation may not work well. Otherwise if only link quality is affected, the link quality measurement can adapt to the changing links. We plan to extend LACA to support mobile and dynamic network conditions.

## 4. Performance Evaluation and Analysis

In this section, we evaluate LACA using simulation experiments. The network is simulated with the TOSSIM [32] simulator. The network contains $10 \times 5$ nodes, of which one node is the sink node and all other 49 nodes are source nodes. Each source node periodically generates the health sensing data and transmits the data to the sink node along a specific path. The pair-wise link quality is shown in Figure 7. The quality of over 80% links is higher than 0.61. At the same time, about 20% links have quality lower than 0.61, on which packets losses are more likely to happen. We tune a number of parameters in the experiment, e.g., the number of paths and available channels, and compare the following key evaluation metrics:

1. Packet delivery ratio (PDR) before deadline. PDR is calculated as the portion of the packets received in the sink node over the total number of data packets generated in the source nodes. It is an end-to-end metric to evaluate the effectiveness of the channel allocation.
2. Probability with insufficient slots. Since colliding links cannot be assigned to the same channel and slot, in some cases there may not be enough slots to be assigned such that the deadline of all paths can be met. This probability is calculated as the fraction of cases with insufficient slots over all allocation cases.
3. The number of transmissions. The number of transmissions denote the total number of transmissions from the time when the first health data is generated to the time when all health data is delivered to the sink node.

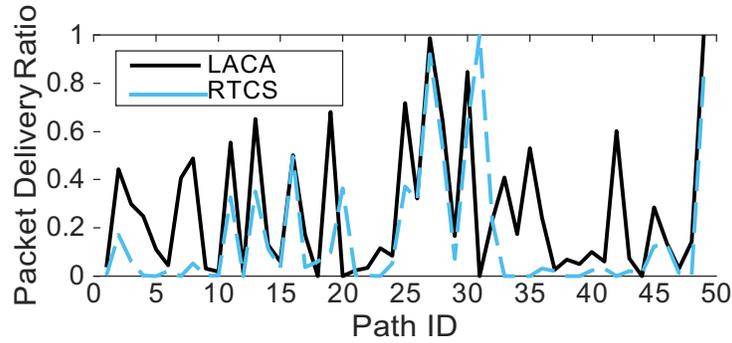

Figure 8: Performance comparison in terms of PDR before deadlines.

4. The number of retransmissions. When packet losses happen, retransmissions are required. Compared to the existing works, retransmissions in LACA can possibly reduce the incurred transmission delay. When there are more retransmissions, the advantage of LACA to the existing works is likely to be larger.

The evaluation and results are shown in the following subsection.

*4.1. Simulation results*

Figure 8 compares the PDR before deadline of LACA and RTCS [17]. It can be inferred that: 1) LACA can greatly improve the PDR before deadline for most paths. The reason is that in LACA, channels/slots with better link quality are prioritized to be assigned. Therefore, the paths in LACA are expected to have better link quality and fewer packet losses. As a result, the PDR is improved significantly. 2) On some paths the PDRs are very close such as path 16, path 27 and path 32. We dig into the experimental traces and identified that the reason is that these paths in LACA have relatively lower priority. Thus the assigned link quality is no better than those in RTCS. The slight improvement of LACA on these paths comes from the additional retransmission slots. 3) The PDRs for some paths are relatively low $\leq 0.5$). The reason is that in our settings, some generated deadlines do not leave any extra slots for retransmissions. As a result, although there can be retransmission slots in LACA, the packet transmissions can be overdue if one packet loss happens in any hop of the path.

Figure 9 shows the fraction of cases with insufficient slots. We repeat the channel allocation 100 times for different number of available channels. The number of cases with insufficient slots are accounted for each different number of available channels. Implications from this figure include:

- LACA has fewer cases with insufficient slots. The reason is that in the path sequence of allocation, paths that have higher collisions with other paths and more urgent deadlines are prioritized, which is expected to distribute the colliding paths into different slots or channels. Differently in the work of RTCS, paths with more urgent deadlines are always prioritized. As a result, some slots/channels which can be used to distribute the colliding paths, are already taken when the colliding paths are to be assigned.

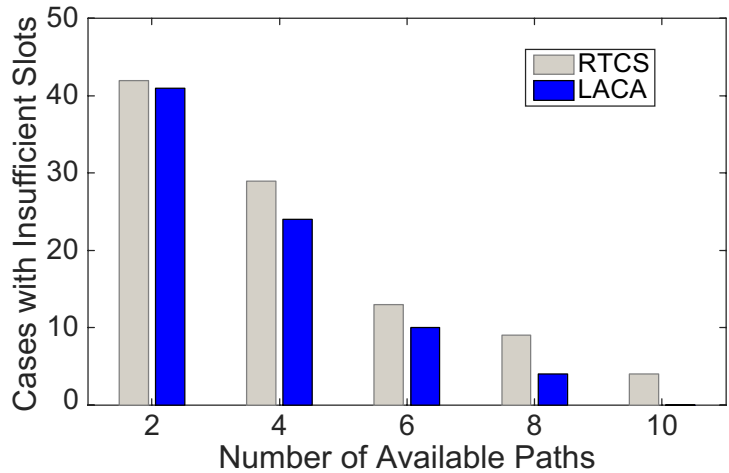

Figure 9: Performance comparison in terms of the number of cases with insufficient slots.

- As the number of available channels increases, the number of cases with insufficient slots decreases for both LACA and RTCS. The reason is that when there are more channel/slot pairs available for the channel allocation, more colliding paths can be fit into the schedule and thus there are fewer cases with insufficient slots.

- As the number of available channels increases, the advantage of LACA over RTCS increases. The reason is that our assigning process first assign the colliding paths while RTCS does not consider the relationship between the assigning order and path collisions. As a result, the number of available slots which can be used to resolve collisions becomes larger.

Figure 10 shows the transmission count comparison of LACA and RTCS. We can infer the following statements: 1) LACA transmits more packets than RTCS for most cases, which seems contradictory to our design that considers link quality. The main reason is that LACA allows for retransmissions and more slots/channels are assigned than RTCS. It is worth noting that these transmissions are also required in RTCS if retransmissions are allows. The difference would be that in RTCS these retransmissions will result in delay of another duty cycle of slots. 2) In some cases, LACA transmits fewer packets than RTCS. This is because LACA considers link quality in channel allocation and the link quality is expected to be better than that with RTCS. As a result, fewer transmissions are required to achieve the same PDR.

Figure 11 shows the retransmission count of LACA. It is worth noting that the cases without retransmissions are not accounted in this figure. Among all the traces, we can see that in 18% cases there is only one retransmission. In 42% cases there are two retransmissions. In 29% cases there are three retransmissions. In 8% cases there are four retransmissions, and in 3% cases there are five retransmissions. To conclude, many links are lossy and retransmissions are largely required. We can also infer that for these cases with retransmissions, the incurred retransmission delay can be significantly reduced since one retransmission in RTCS incurs an delay of a full duty cycle of slots while one retransmission in LACA only incurs the delay of several slots fewer than a duty cycle.

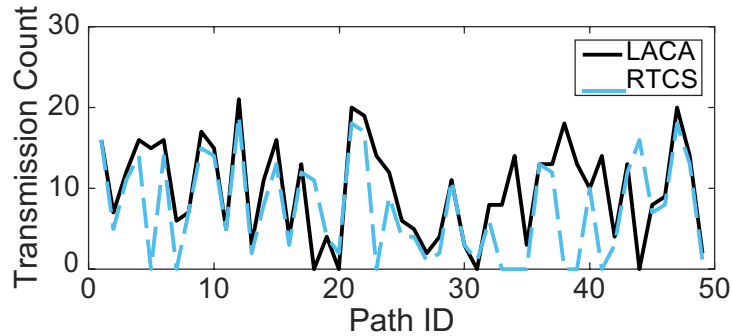

Figure 10: Performance comparison in terms of the number of transmissions.

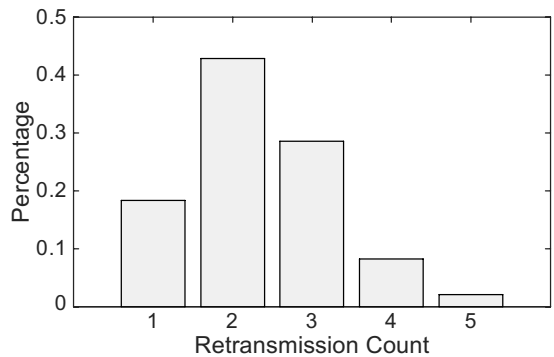

Figure 11: Performance comparison in terms of retransmission count.

Next we compare the resource utilization of LACA and RTCS. The metric of resource utilization is defined as the fraction of slots that are actually utilized for transmissions/retransmissions over the total number of slots in an entire schedule. A larger resource utilization means more spare slots are used for retransmissions since the number of slots for first-round transmissions are the same for LACA and RTCS (which equals to the number of links in the network topology). Figure 12 shows the result. The resource utilization of LACA is much higher than RTCS work. The reason is that we intentionally assign spare slots for possible retransmissions. These slots will be used if there are packet losses. It is worth noting that these packets will not incur any energy consumption if there are no packet losses. It can also be inferred that without these retransmission slots, the PDR of LACA will also decrease.

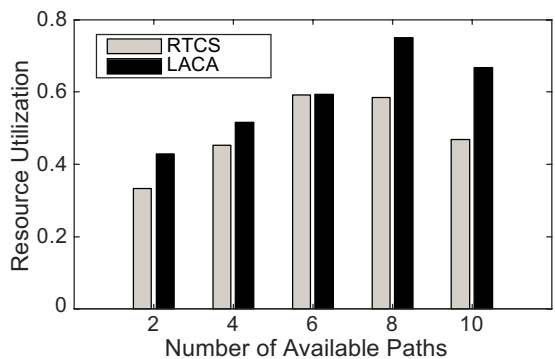

Figure 12: Performance comparison in terms of resource utilization.

## 5. Conclusion

In this paper, we present a novel channel allocation scheme in body sensor networks (call LACA), which jointly considers both link quality, collisions and deadlines. Compared to the existing works, LACA prioritizes the paths with more urgent deadlines and heavier collisions in the channel allocation process and therefore more paths can be fit into the schedule before deadline. For each path link allocation, the channel/slot with better link quality is preferred such that there are fewer packet losses. Moreover, the spare slots after the first round allocation are allowed to be used for possible retransmissions. Extensive simulation experiments are conducted, showing that LACA greatly improves the packet delivery ratio (PDR) before deadline compared to the state-of-the-art works.

**Acknowledgment**

This work was supported by the Fundamental Research Funds for the Central Universities (No. ZYGX2016KYQD098), the National Natural Science Foundation of China (No. 61602095) and the EU FP7 CLIMBER project under Grant Agreement No. PIRSES-GA-2012-318939.